\documentclass[runningheads]{llncs}
\usepackage{graphicx}
\usepackage{xcolor}
\usepackage{float}
\usepackage{multirow}
\usepackage{authblk}
\begin{document}
\title{An Empirical Study on the Fairness of Foundation Models for Multi-Organ Image Segmentation}
\titlerunning{Fairness of Foundation Models in Multi-organ Image Segmentation}
%
\author{Qing Li$^\#$
\inst{1} \and Yizhe Zhang$^\#$ \inst{2} \and Yan Li \inst{3} \and Jun Lyu \inst{4} \and Meng Liu \inst{5} \and Longyu Sun \inst{1} \and Mengting Sun \inst{1} \and Qirong Li\inst{5} \and Wenyue Mao\inst{6} \and Xinran Wu\inst{6} \and Yajing Zhang\inst{7} \and Yinghua Chu\inst{8} \and Shuo Wang$^*$\inst{9}
 \and Chengyan Wang$^*$\inst{1}}
\authorrunning{Q. Li et al.}


\institute{
\textsuperscript{1}Human Phenome Institute, Fudan University, Shanghai, China\\
\inst{2}School of Computer Science and Engineering, Nanjing University of Science and Technology, Nanjing, Jiangsu China\\
\inst{3}Department of Radiology, Ruijin Hospital, Shanghai Jiao Tong University School of Medicine, Shanghai, China\\
\inst{4}Department of Psychiatry, Brigham and Women's Hospital, Harvard Medical School, Boston, Massachusetts, US\\
\inst{5}School of Computer Science, Fudan University, Shanghai, China\\
\inst{6}Institute of Science and Technology for Brain-Inspired Intelligence, Fudan University, Shanghai, China \\
\inst{7}MR Business Unit, Philips Healthcare Suzhou, China\\
\inst{8}Simens Healthineers Ltd., China\\
\inst{9}Digital Medical Research Center, School of Basic Medical Sciences, Fudan University, Shanghai, China\\
}

\maketitle              
\def\thefootnote{\#}\footnotetext{Equal contribution.}
\def\thefootnote{*}\footnotetext{Corresponding authors: wangcy@fudan.edu.cn and shuowang@fudan.edu.cn }
\begin{abstract}
The segmentation foundation model, e.g., Segment Anything Model (SAM), has attracted increasing interest in the medical image community. Early pioneering studies primarily concentrated on assessing and improving SAM's performance from the perspectives of overall accuracy and efficiency, yet little attention was given to the fairness considerations. This oversight raises questions about the potential for performance biases that could mirror those found in task-specific deep learning models like nnU-Net. In this paper, we explored the fairness dilemma concerning large segmentation foundation models. We prospectively curate a benchmark dataset of 3D MRI and CT scans of the organs including liver, kidney, spleen, lung and aorta from a total of 1056 healthy subjects with expert segmentations. Crucially, we document demographic details such as gender, age, and body mass index (BMI) for each subject to facilitate a nuanced fairness analysis. We test state-of-the-art foundation models for medical image segmentation, including the original SAM, medical SAM and SAT models, to evaluate segmentation efficacy across different demographic groups and identify disparities. Our comprehensive analysis, which accounts for various confounding factors, reveals significant fairness concerns within these foundational models. Moreover, our findings highlight not only disparities in overall segmentation metrics, such as the Dice Similarity Coefficient but also significant variations in the spatial distribution of segmentation errors, offering empirical evidence of the nuanced challenges in ensuring fairness in medical image segmentation.

\end{abstract}
\section{Introduction}
Since the introduction of the Segment Anything Model (SAM), early attempts have been made to evaluate~\cite{huang2024segment}, adapt~\cite{wu2023medical}, and utilize~\cite{zhang2023input} SAM for medical image segmentation. Studies have found that the direct application of SAM to medical images leads to unsatisfactory segmentation results~\cite{huang2024segment,he2023accuracy}. It has been understood that SAM was originally trained on natural scene images, therefore a foundation model trained or fine-tuned for medical images is beneficial. Motivated by this, Ma et al.\cite{ma2024segment} utilized a collection of public datasets to train a medical SAM tailored specifically for medical image segmentation. In a similar spirit, Cheng et al.\cite{cheng2023sammed2d} built a 2D medical SAM, while Wang et al.~\cite{wang2023sam} constructed a 3D medical SAM, relying on a combination of public datasets and their private datasets. The foundation models, particularly SAM, represent an emerging focal point within the realm of Medical AI research. Most existing studies have focused on investigating the overall segmentation accuracy and efficiency of SAM on medical images, with scant attention given to the fairness considerations of these emerging large foundation models.

The issue of fairness in medical image analysis has sparked significant attention, with numerous studies shedding light on the subject (e.g, ~\cite{gaggion2023unsupervised,afzal2023towards,seyyed2021underdiagnosis,chen2023algorithmic,xu2023fairness}. The fairness issue stems from inherent inductive biases and distributional discrepancies between training and evaluation datasets. Demographic disparities among patient groups can lead to variations in model performance, particularly concerning organ characteristics influenced by factors such as gender and age. This can introduce bias into segmentation outcomes, affecting the fairness of the models. Despite efforts aimed at enhancing model generalization through techniques such as combining datasets~\cite{ma2024segment} and federated learning~\cite{yoon2023privacy}, yet fairness challenges remain formidable. This persistence of fairness issues prompts a comparison with the well-documented biases of task-specific segmentation models, such as U-Net, raising questions about the efficacy of foundation models trained on extensive image datasets in addressing these biases. The ongoing assessment of these challenges is crucial in our quest to surmount the inherent limitations of large-scale medical models, ensuring equitable and unbiased medical diagnostics. 

In this paper, we study the fairness of the original SAM~\cite{kirillov2023segment}, Medical SAM~\cite{ma2024segment}, and recently developed SAT models~\cite{zhao2023model} in segmenting multiple organs (including the liver, kidneys, and spleen in MRI, as well as the lungs and aorta in CT scans). Gender, age and BMI are considered as sensitive attributions for the fairness study. In summary, this work contributes in the following three aspects. 
\begin{itemize}
\item We conducted a detailed and comprehensive comparison of the segmentation performance of emerging large segmentation foundation models, including the original SAM, Medical SAM, and SAT, for medical images.

\item Our study of fairness on the segmentation foundation models addresses multiple body parts, i.e., liver, kidney, spleen, lung, aorta, concerning sensitive attributes, i.e., gender, age and BMI. For the first time, we study BMI attribute in the fairness problem in the context of medical image segmentation.

\item We delved deeper into studying fairness at the organ sub-regions and spatial aspects. Experiments, accompanied by visualization results, unveiled new insights and issues regarding fairness and performance variations at the organ sub-region level.

\end{itemize}



\begin{figure}[t]
   \centering
   \includegraphics[width=1.0\linewidth]{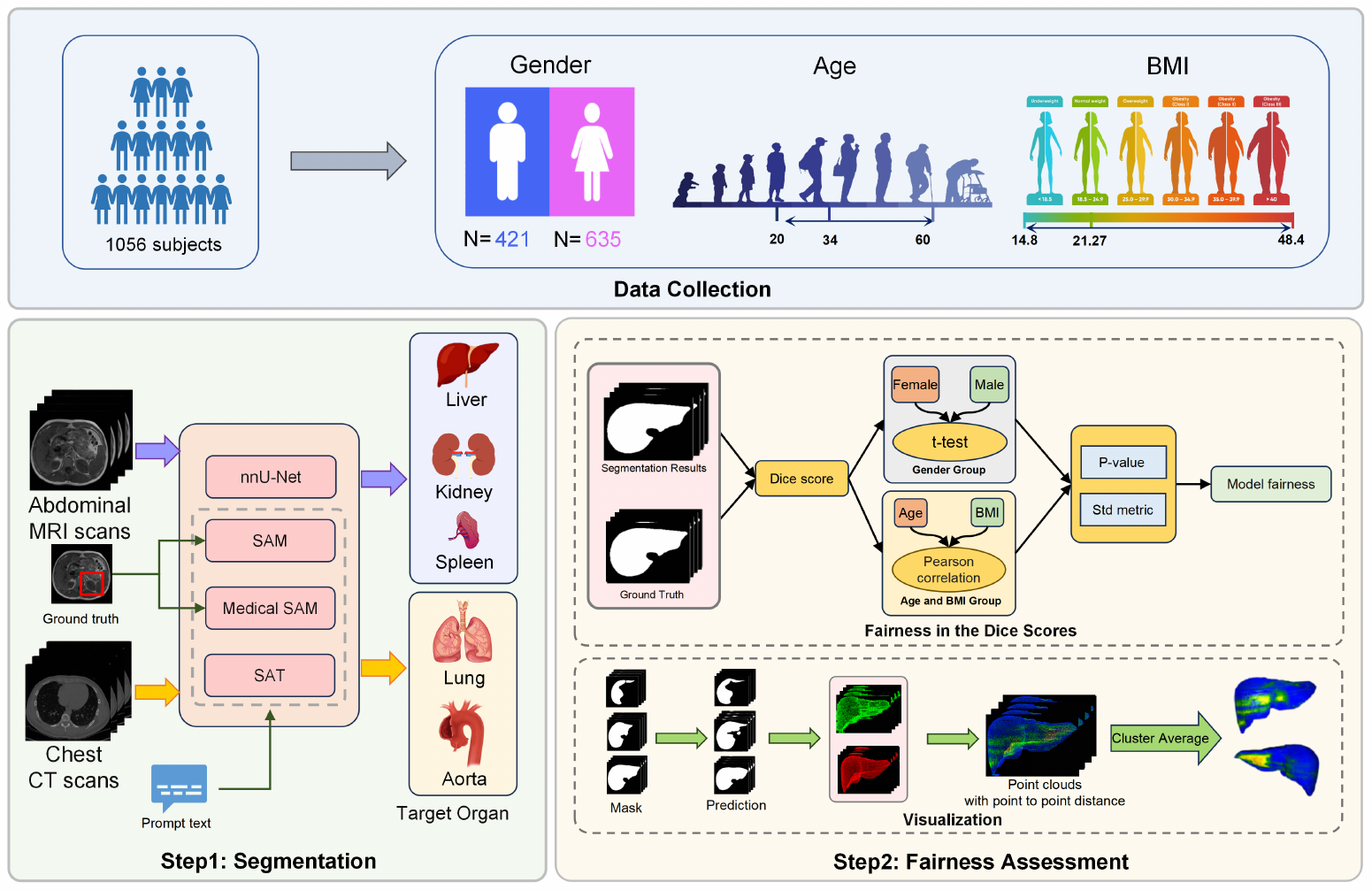}
   \caption{Overview: fairness study on segment foundation models for multi-organ images.}
   \label{fig:enter-label}
\end{figure}

\section{Setups}
\subsection{Data Collection}\label{sec:DC}
This study employed local abdominal MRI scans and chest CT scans as datasets to investigate the fairness of segmentation performance. The abdominal scans, including the liver, spleen, and kidneys, utilized the two-echo mDIXON-TSE technology\cite{dixon1984simple} to ensure high signal-to-noise ratios and excellent image contrast. Imaging was conducted on the axial plane using the breath-hold technique. The field of view was selected to encompass the liver, spleen, pancreas, and kidneys. Key imaging parameters included field of view of 350×350 mm², spatial resolution of 1.5×1.6 mm², a total of around 100 slices, slice thickness of 5.0 mm, repetition time of 3.4 ms, echo times (TR/TE) of 1.31/2.20 ms, flip angle of 10 degrees, and a sensitivity encoding (SENSE) factor of 2.0.

Chest CT scans for the lung and aorta were conducted with the following parameters: field of view of 350×200 mm², spatial resolution of 0.68×0.68 mm², tube voltage of 100 kV, tube current of 134 mAs, a total of around 400 slices, and slice thickness of 1.0 mm. The breath-holding protocol was consistently applied to ensure image stability. Image reconstruction was facilitated by fast cone-beam filtered back projection, producing images with a reconstructed thickness of 1.0 mm. The lung images were configured with a window width of 1,000 HU and a window level of -650 HU, while the mediastinum was adjusted to a window width of 350 HU and a window level of 40 HU.

The collected dataset comprises abdominal MRI and thoracic CT scan images from 1,056 volunteers, encompassing 421 males and 635 females aged between 20 and 60 years, with BMIs  ranging from 14.8 to 48.4. The study obtained approval from the ethics committee of the local hospital, and the data annotation was meticulously carried out by two technicians, each boasting over five years of professional experience. Furthermore, the gender, age, and BMI information of all volunteers were provided and authorized for use by our local institution.

\subsection{Segmentation Foundation Models under Investigation}
In this study, we investigate several popular segmentation foundation models, namely the original SAM~\cite{kirillov2023segment}, Medical SAM~\cite{ma2024segment}, and SAT~\cite{zhao2023model}, for the segmentation of multiple organs, including the liver, spleen, kidneys, lung, and aorta, from MRI scans of the abdomen and CT scans of the chest. To compare and contrast their performance, we further train a nnU-Net and evaluate it using the same test samples employed for evaluating the foundational models.

Trained SAM, Medical SAM, and SAT were downloaded from their publishers, and we test each model using the recommended settings. It's noted that SAM and Medical SAM require additional prompts for generating segmentation masks. Thus, we adopt a common practice for testing SAM by utilizing the ground truth to derive bounding boxes/center points of each object to construct the prompts. As for SAT, which requires text input, we simply employ the name of the object of interest (e.g., liver) along with an image to prompt SAT. The nnU-Net, on the other hand, is trained on a dataset collected following the same pipeline but during a distinct temporal interval. It is reasonable to assume that the data utilized in training nnU-Net is similar to the test samples, but not precisely from the same data distribution.



\subsection{Assessing Fairness}\label{sec:AF}

\textbf{Quantitative Assessment.} The Dice score serves as the primary metric for evaluating segmentation performance. 
These scores for individual test samples are grouped according to specific attributes, such as gender, age, and BMI.
We first examine whether a significant difference exists between two groups of Dice scores. The t-test is utilized to quantify the discrepancy between the gender (male and female) while Pearson correlation analysis is employed for age and BMI. 
The p-values obtained from the above quantification reflect whether different groups exhibit similar scores in distribution, thus indicating fairness.
 Following ~\cite{wang2020mitigating}, we further quantify the inter-group performance disparities, namely how much the mean Dice score of each group spread by calculating the standard deviation of the group means (Std-GM). A high Std-GM suggests a more unequal segmentation performance with varying attributes. 


\noindent\textbf{Sub-region and Spatial Fairness.} In addition, we assess fairness in segmentation outputs from a spatial perspective. Using the liver as an example, for each liver pixel in the ground truth (GT) of a test sample, we identify the closest liver pixel in the corresponding segmentation map based on Euclidean distance, recording the distance value. This process is repeated for all liver pixels in the GT to generate a distance map, representing the alignment of the segmentation map with the GT map; larger distances indicate unsatisfactory segmentation performance. For a group of patients (e.g., male patients) comprising $m$ test samples, we repeat the above process for each test sample to obtain $m$ distance maps. These maps are then combined using K-Means technique  to generate a final distance map for the group. The resulting distance map can be visualized as a heat map and used to qualitatively compare sub-regions and spatial fairness between groups. More details of the algorithm can be found in the supplementary materials.

\section{Experiments and Results}

\begin{table*}[t]
\centering
\scriptsize
\caption{Segmentation performance in groups specified by the gender attribute. }
\begin{tabular}{|c|c|c|c|c|c|c|}
\hline

Models & Gender     & Liver     & Kidney     & Spleen     &  Lung & Aorta      \\ \cline{1-7} 
\multirow{4}{*}{nnU-Net}       & Male       & 0.968±0.041 & 0.943±0.089 & 0.899±0.175 & 0.992±0.055 & 0.860±0.252          \\ \cline{2-7} 
                                & Female       & 0.961±0.056 & 0.961±0.062 & 0.935±0.090 & 0.995±0.001 & 0.847±0.257           \\ \cline{2-7} 
                                & Std-GM$\uparrow$       & 0.0046        & 0.0129          & \underline{0.0256}         & 0.0020        & 0.0088            \\ \cline{2-7}
                                & P-value$\downarrow$      & 0.0667    & 0.0022    & 0.0012	&0.3414	& 0.5315   \\ \hline
\multirow{4}{*}{SAM}         & Male        & 0.900±0.043 & 0.896±0.320 & 0.851±0.069 & 0.9594±0.009 & 0.7871±0.033           \\ \cline{2-7} 
                                & Female       & 0.856±0.059 & 0.886±0.021 & 0.820±0.064 & 0.943±0.115 & 0.769±0.031           \\ \cline{2-7} 
                                 & Std-GM$\uparrow$      & \underline{0.0310}	& 0.0067 &	0.0224	& 0.0114	& 0.0130            \\ \cline{2-7}
                                & P-value$\downarrow$     &\textbf{\textless0.0001}	&\textbf{\textless0.0001}	&\textbf{\textless0.0001}	&0.0048	&\textbf{\textless0.0001}                               \\ \hline
\multirow{4}{*}{Medical SAM}         & Male        & 0.795±0.061 & 0.883±0.025 & 0.819±0.060 & 0.901±0.014 & 0.642±0.068           \\ \cline{2-7} 
                                & Female       & 0.772±0.073 & 0.878±0.014 & 0.808±0.053 & 0.865±0.107 & 0.676±0.063           \\ \cline{2-7} 
                                  & Std-GM$\uparrow$       &0.0161	&0.0038	&0.0083	&\underline{0.0253}	&\underline{0.0237}            \\ \cline{2-7}
                                & P-value$\downarrow$     & \textbf{\textless0.0001}	&0.0008	&0.0064	& \textbf{\textless0.0001}	&\textbf{\textless0.0001}   \\ \hline
\multirow{4}{*}{SAT}        & Male       & 0.875±0.061 & 0.567±0.064 & 0.712±0.258 & 0.987±0.001 & 0.641±0.191           \\ \cline{2-7} 
                                & Female       & 0.880±0.059 & 0.546±0.089 & 0.677±0.262 & 0.987±0.002 & 0.619±0.198           \\ \cline{2-7} 
                                & Std-GM$\uparrow$       &0.0037	& \underline{0.0148}	& 0.0247	& 0.0003	& 0.0157            \\ \cline{2-7}
                                & P-value$\downarrow$    &0.2409	& 0.0002 &	0.0746	& 0.0001 & 	0.1504                                   \\ \hline

\hline
\end{tabular}
\label{table1}
\end{table*}

\begin{table*}[t]
\centering
\scriptsize
\caption{Segmentation performance in groups specified by the age attribute.}
\begin{tabular}{|c|c|c|c|c|c|c|}
\hline

Models & Age     & Liver     & Kidney     & Spleen     &  Lung & Aorta      \\ \cline{1-7} 
\multirow{6}{*}{nnU-Net}       & 20\texttildelow30       & 0.962±0.050& 0.953±0.083& 0.924±0.109& 0.992±0.054&0.819±0.284         \\ \cline{2-7} 

                                & 30\texttildelow40       & 0.966±0.037& 0.958±0.074& 0.917±0.147& 0.995±0.001&0.864±0.243         \\ \cline{2-7} 
                                & 40\texttildelow50       & 0.963±0.076& 0.948±0.065& 0.919±0.137& 0.995±0.001&0.903±0.195          \\ \cline{2-7} 
                                & 50\texttildelow60       & 0.966±0.038& 0.955±0.057& 0.916±0.159& 0.995±0.001&0.873±0.228        \\ \cline{2-7} 
                                & Std-GM$\uparrow$      & 0.0020 &	0.0040	& 0.0035	& 0.0014	& 0.0348           \\ \cline{2-7} 
                                & P-value$\downarrow$ & 0.4731	& 0.7262	& 0.6046	& 0.6739 & 	0.0156\\ \hline
\multirow{6}{*}{SAM}       & 20\texttildelow30       & 0.853±0.059& 0.883±0.026& 0.805±0.0680& 0.951±0.079& 0.760±0.028       \\ \cline{2-7} 

                                & 30\texttildelow40       & 0.884±0.052& 0.895±0.023& 0.840±0.065& 0.959±0.009& 0.774±0.031         \\ \cline{2-7} 
                                & 40\texttildelow50       & 0.886±0.056& 0.896±0.030& 0.852±0.062& 0.950±0.090& 0.791±0.025          \\ \cline{2-7} 
                                & 50\texttildelow60       & 0.900±0.041& 0.897±0.019& 0.873±0.043& 0.933±0.154& 0.806±0.026       \\ \cline{2-7} 
                                & Std-GM$\uparrow$    &  \underline{0.0198}&	0.0063&	\underline{0.0283}&	\underline{0.0108}&	0.0198           \\ \cline{2-7} 
                                & P-value$\downarrow$ & \textbf{\textless0.0001} &	\textbf{\textless0.0001} &	\textbf{\textless0.0001}&	0.0692 &	\textbf{\textless0.0001}\\ \hline
\multirow{6}{*}{Medical SAM}       & 20\texttildelow30       & 0.764±0.069& 0.876±0.019& 0.796±0.058& 0.880±0.075&0.669±0.067        \\ \cline{2-7} 

                                & 30\texttildelow40       & 0.795±0.061& 0.883±0.017& 0.815±0.052& 0.886±0.020&0.662±0.065         \\ \cline{2-7} 
                                & 40\texttildelow50       & 0.791±0.072& 0.885±0.024& 0.819±0.055& 0.878±0.086&0.656±0.064          \\ \cline{2-7} 
                                & 50\texttildelow60       & 0.797±0.069& 0.882±0.017& 0.844±0.041& 0.868±0.144&0.655±0.069       \\ \cline{2-7} 
                                & Std-GM$\uparrow$      &0.0154	& 0.0038	& 0.0195	& 0.0073	& 0.0061          \\ \cline{2-7} 
                                & P-value$\downarrow$ & \underline{\textless0.0001} &	0.0002 &	\underline{\textless0.0001} &	0.1438	& 0.1545\\ \hline


\multirow{6}{*}{SAT}       & 20\texttildelow30       & 0.882±0.049& 0.545±0.085& 0.668±0.277& 0.987±0.001& 0.582±0.210        \\ \cline{2-7} 
                                & 30\texttildelow40       & 0.882±0.051& 0.564±0.068& 0.694±0.261& 0.987±0.002& 0.635±0.185        \\ \cline{2-7} 
                                & 40\texttildelow50       & 0.871±0.094& 0.557±0.093& 0.726±0.218& 0.987±0.002& 0.693±0.143          \\ \cline{2-7} 
                                & 50\texttildelow60       & 0.871±0.051& 0.560±0.065& 0.711±0.250& 0.987±0.002& 0.678±0.184        \\ \cline{2-7} 
                                & Std-GM$\uparrow$     & 0.0065	& \underline{0.0081}	& 0.0248	& 0.0001	& \underline{0.0498}           \\ \cline{2-7} 
                                & P-value$\downarrow$ & 0.0208	& 0.0432&	0.0522&	0.9736&	\textbf{{\textless0.0001}}\\ \hline

\hline
\end{tabular}
\label{table2}
\end{table*}

\subsection{Fairness over Individual Attributes}
We apply the original SAM, Medical SAM, and SAT models to the data we collected (see Sec.~\ref{sec:DC}) using the inference pipeline suggested by the model publishers. For the 1056 patients, we obtain the segmentation results for each organ.

We split the lists according to the attributes under investigation, namely age, gender, and BMI. Taking gender as an example, for each organ type, we divide the corresponding 1056 Dice scores into two groups: the male group and the female group. Subsequently, we calculate the average Dice score within each group and report both the averages and standard deviations. In addition, we perform statistical test and compute the Std-GM (see Sec.~\ref{sec:AF}) for the male and female groups. We perform the above process for each attribute (gender, age, BMI) and report the performances in Tables~\ref{table1}, \ref{table2}, and \ref{table3} for the gender, age, and BMI attributes, respectively.

\begin{table*}[t]
\centering
\scriptsize
\caption{Segmentation performance in groups specified by the BMI attribute.}
\begin{tabular}{|c|c|c|c|c|c|c|}
\hline
Models & BMI     & Liver     & Kidney     & Spleen     &  Lung & Aorta      \\ \cline{1-7} 
\multirow{5}{*}{nnU-Net}       & Underweight       & 0.948±0.035& 0.969±0.027& 0.930±0.049& 0.995±0.001&0.770±0.287        \\ \cline{2-7} 

                                & Healthy       & 0.965±0.051& 0.963±0.053& 0.930±0.106& 0.995±0.001&0.847±0.260         \\ \cline{2-7} 
                                & Overweight       & 0.964±0.052& 0.934±0.105& 0.901±0.177& 0.991±0.062&0.876±0.235          \\ \cline{2-7} 

                                & Std-GM$\uparrow$      & 0.0094	& \underline{0.0187}	& 0.0168	& 0.0026	& \underline{0.5490}           \\ \cline{2-7} 
                                & P-value$\downarrow$    & 0.5793	&\textbf{{\textless0.0001}} &	0.0008 &	0.3708 &	0.1241                   \\ \hline
\multirow{5}{*}{SAM}       & Underweight       & 0.803±0.069& 0.870±0.022& 0.778±0.065& 0.959±0.0080& 0.777±0.036       \\ \cline{2-7} 

                                & Healthy       & 0.859±0.054& 0.885±0.023& 0.817±0.064& 0.951±0.081& 0.771±0.031        \\ \cline{2-7} 
                                & Overweight       & 0.911±0.037& 0.902±0.027& 0.870±0.058& 0.946±0.113& 0.784±0.035         \\ \cline{2-7} 
                                & Std-GM$\uparrow$      &0.0537	& 0.0159	& 0.0462	& \underline{0.0069}	& 0.0068           \\ \cline{2-7} 
                                & P-value$\downarrow$    &\textbf{\textless0.0001}	&\textbf{\textless0.0001}	&\textbf{\textless0.0001}	& 0.0847	& 0.0003                 \\ \hline
\multirow{5}{*}{Medical SAM}       & Underweight      & 0.699±0.095& 0.874±0.012& 0.797±0.045& 0.884±0.019&0.679±0.074        \\ \cline{2-7} 

                                & Healthy       & 0.774±0.063& 0.876±0.017& 0.802±0.056& 0.878±0.077&0.670±0.061         \\ \cline{2-7} 
                                & Overweight      & 0.808±0.064& 0.888±0.023& 0.835±0.052& 0.881±0.106&0.645±0.072          \\ \cline{2-7} 
                                & Std-GM$\uparrow$      & \underline{0.0555}	& 0.0074	& 0.0204	& 0.0030	& 0.0175           \\ \cline{2-7} 
                                & P-value$\downarrow$    & \textbf{\textless0.0001}	& \textbf{\textless0.0001}	& \textbf{\textless0.0001}	& 0.3865	& 0.0001                   \\ \hline

\multirow{5}{*}{SAT}       & Underweight      & 0.856±0.064& 0.545±0.109& 0.612±0.289& 0.987±0.001& 0.547±0.215        \\ \cline{2-7} 
                                & Healthy      & 0.887±0.055& 0.551±0.083& 0.676±0.272& 0.987±0.001& 0.622±0.200        \\ \cline{2-7} 
                                & Overweight       & 0.865±0.066& 0.562±0.069& 0.730±0.227& 0.987±0.002& 0.652±0.181          \\ \cline{2-7} 
                                & Std-GM$\uparrow$      & 0.0160	& 0.0086	& \underline{0.0589}	&0.0004	& 0.0543           \\ \cline{2-7} 
                             & P-value$\downarrow$    &0.0001	&0.4970	& 0.0060	& \textbf{\textless0.0001}	& 0.0453                   \\ \hline

\hline
\end{tabular}
\label{table3}
\end{table*}

In Table~\ref{table1}, we observe moderate to severe fairness problems for SAM and Medical SAM in organ segmentation across genders. SAM yields a more unfair segmentation performance. Medical SAM, trained using a large collection of medical images, exhibits fewer unfair performance issues (according to p-value) than the original SAM. Among the three tested foundation models, SAT demonstrates the most fair segmentation performance. On the other hand, since only texts were used as prompts, the actual segmentation performance from SAT is not on par with the other two bounding-box prompted SAMs. The nnU-Net, as a reference model, delivers overall the best segmentation performance. This is partially due to nnU-Net being an in-house trained model, with curated training samples more closely related to the test samples than the training samples used in training the generalists (e.g., SAM).

We observe a similar phenomenon where SAM produces the worst results in fairness, while Medical SAM, being a fairer model (than SAM), still does not match the fairness of the specialist (nnU-Net) in Table~\ref{table2}. Again, SAT yields the worst overall segmentation results due to the lack of more explicit prompts (e.g., bounding-box), despite being a fairer model than the SAM counterparts.

With BMI as a sensitive attribute, more interesting observations can be found in Table~\ref{table3}. Firstly, nnU-Net, for the first time, exhibits unfair segmentation performance, particularly in the kidney segmentation task. Upon closer inspection, we find that the overweight group suffers greatly in segmentation performance when nnU-Net segmenting those kidney areas in the images. Both Medical SAM and SAM perform unfairly for the liver, kidney, and spleen classes. Upon inspecting the Std-GM, we note that for the liver, SAM and Medical SAM exhibit a similar level of unfairness (0.053 $\approx$ 0.055) in segmentation performance. For the kidney and spleen, although Medical SAM demonstrates unfair performance (indicated by the p-values), the severity of unfairness is less pronounced than that of SAM (0.0074 $<$ 0.0159 for the kidney and 0.0204 $<$ 0.0462 for the spleen). 

\textbf{In summary}, after inspecting Tables~\ref{table1}, \ref{table2}, and \ref{table3}, we highlight the following observations. (1) In general, Medical SAM delivers fairer segmentation performance than the original SAM but exhibits worse overall accuracy. (2) Among the three foundation models tested, SAT yields the best results in terms of performance fairness but exhibits the worst overall accuracy. (3) nnU-Net, since it is trained with in-house data that is better curated than the collection of public datasets used in training the generalists (e.g., Medical SAM), provides the best results in terms of both fairness considerations and overall segmentation accuracy. (4) The organs of lung and Aorta receive fairer segmentation treatment across models comparing other tested organs.

\begin{figure}[t]
   \centering
   \includegraphics[width=1.0\linewidth]{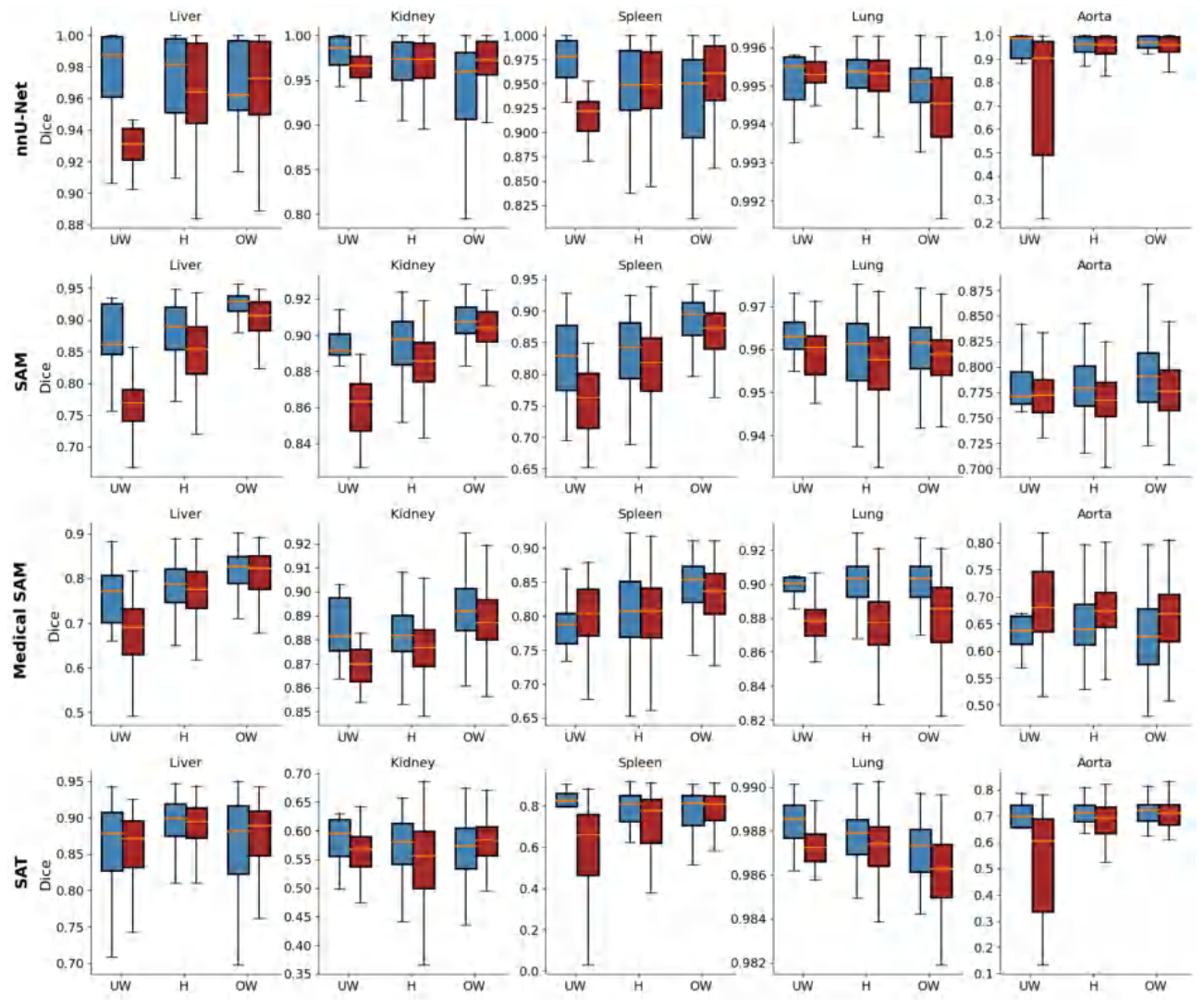}
    \caption{Segmentation performance (means and standard deviations) for subject groups specified by combinations of gender and BMI attributes. Blue box: male. Red box: female. UW: Underweight, H: Healthy, OW: Overweight.}
   \label{fig:box-plots}
\end{figure}

\subsection{Fairness over Joint Attributes}
Furthermore, we showcase the segmentation performance discrepancy between male and female groups under different BMI levels and highlight some intriguing findings (in Fig.~\ref{fig:box-plots}). First, for the well-trained nnU-Net, in liver segmentation, females with underweight BMI levels receive significantly worse segmentation results than their male counterparts. This phenomenon also exists for a range of models on different organs. For instance, nnU-Net and Medical SAM on the liver, SAM, Medical SAM, and SAM on the kidney, nnU-Net, SAM, and SAT on the spleen, Medical SAM, and SAT on the lung, and nnU-Net and SAT on the aorta. This finding is worth further investigation. The other reports of fairness over joint attributes can be found in the figures in supplementary materials. 

\subsection{Fairness in Sub-regions of Organs}
\begin{figure}[t]
   \centering
   \includegraphics[width=1.0 \linewidth]{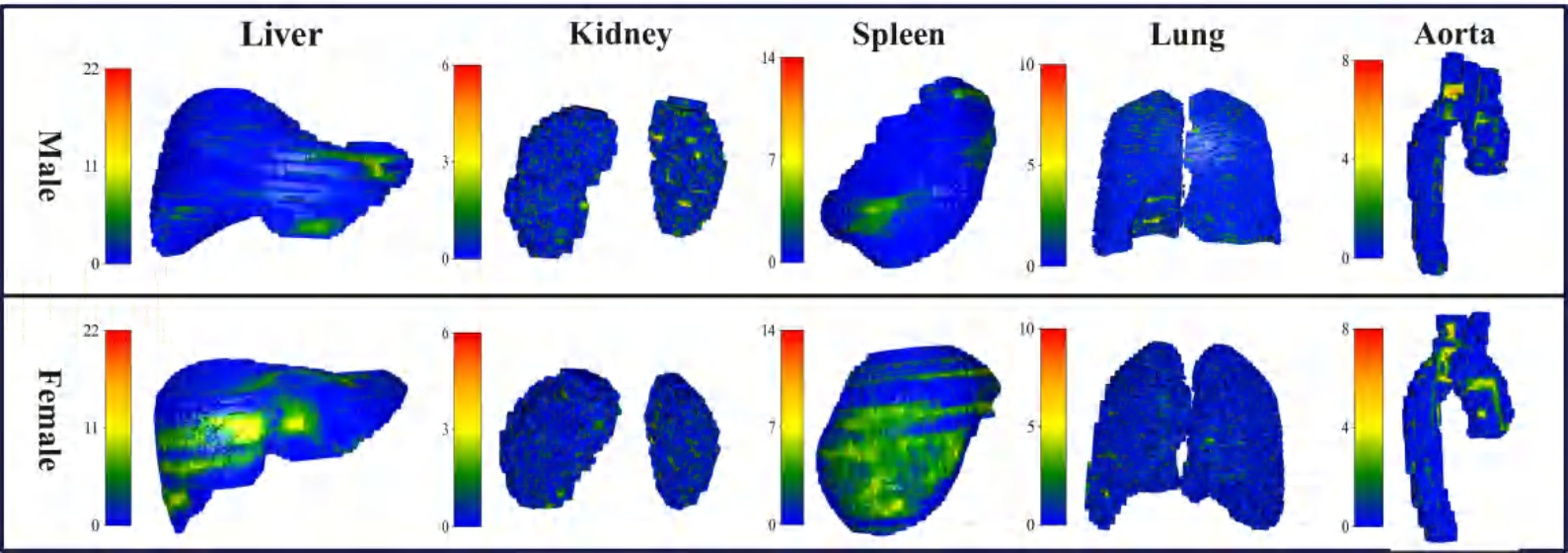}
   \caption{Visualization of segmentation errors in sub-regions (mean distances to GT).}
   \label{fig:sub-regions}
\end{figure}

We further provide visualizations of the segmentation errors for the male and female groups across all the organs studied in Fig.~\ref{fig:sub-regions} by Medical SAM. It is visually evident that there exists a clear bias/unfairness in segmentation errors across sub-regions in the segmentation region for certain organs. It is notable that, in the case of liver segmentation, females are more prone to experiencing errors in the right lobe of liver. Likewise, when it comes to spleen, images from females are also more likely to exhibit errors in the forehead regions.

\section{Conclusion}
This study conducted a comprehensive study on the fairness performance of emerging segmentation foundation models for medical image segmentation. Our study revealed the existence of fairness issues and their varying degrees in the original SAM, Medical SAM, and SAT models. Compared to an in-house trained specialist model, nnU-Net, these segmentation foundational models demonstrated significant fairness problems. Our study underscores the need for increased attention and effort in addressing fairness issues during the development, comparison, utilization and quality control of foundational models in medical applications.


\bibliographystyle{plain}
\bibliography{reference}

\end{document}